\documentclass[pdflatex,sn-mathphys-num]{sn-jnl}

\usepackage{graphicx}%
\usepackage{multirow}%
\usepackage{amsmath,amssymb,amsfonts}%
\usepackage{amsthm}%
\usepackage{mathrsfs}%
\usepackage[title]{appendix}%
\usepackage{xcolor}%
\usepackage{textcomp}%
\usepackage{manyfoot}%
\usepackage{booktabs}%
\usepackage{algorithm}%
\usepackage{algorithmicx}%
\usepackage{algpseudocode}%
\usepackage{listings}%
\usepackage{nameref}
\usepackage{tipa}
\usepackage{multibib}

\ifx \bisbn   \undefined \def \bisbn  #1{ISBN #1}\fi
\ifx \binits  \undefined \def \binits#1{#1}\fi
\ifx \bauthor  \undefined \def \bauthor#1{#1}\fi
\ifx \batitle  \undefined \def \batitle#1{#1}\fi
\ifx \bjtitle  \undefined \def \bjtitle#1{#1}\fi
\providecommand{\aap}{A\&A}%
\providecommand{\aj}{AJ}%
\providecommand{\apj}{ApJ}%
\providecommand{\apjl}{ApJ}%
\providecommand{\araa}{ARA\&A}%
\providecommand{\mnras}{MNRAS}%
\providecommand{\nar}{New Astron. Rev.}%
\providecommand{\nat}{Nature}%
\ifx \bvolume  \undefined \def \bvolume#1{\textbf{#1}}\fi
\ifx \byear  \undefined \def \byear#1{#1}\fi
\ifx \bissue  \undefined \def \bissue#1{#1}\fi
\ifx \bfpage  \undefined \def \bfpage#1{#1}\fi
\ifx \blpage  \undefined \def \blpage #1{#1}\fi
\ifx \burl  \undefined \def \burl#1{\textsf{#1}}\fi
\ifx \doiurl  \undefined \def \doiurl#1{\url{https://doi.org/#1}}\fi
\ifx \betal  \undefined \def \betal{\textit{et al.}}\fi
\ifx \binstitute  \undefined \def \binstitute#1{#1}\fi
\ifx \binstitutionaled  \undefined \def \binstitutionaled#1{#1}\fi
\ifx \bctitle  \undefined \def \bctitle#1{#1}\fi
\ifx \beditor  \undefined \def \beditor#1{#1}\fi
\ifx \bpublisher  \undefined \def \bpublisher#1{#1}\fi
\ifx \bbtitle  \undefined \def \bbtitle#1{#1}\fi
\ifx \bedition  \undefined \def \bedition#1{#1}\fi
\ifx \bseriesno  \undefined \def \bseriesno#1{#1}\fi
\ifx \blocation  \undefined \def \blocation#1{#1}\fi
\ifx \bsertitle  \undefined \def \bsertitle#1{#1}\fi
\ifx \bsnm \undefined \def \bsnm#1{#1}\fi
\ifx \bsuffix \undefined \def \bsuffix#1{#1}\fi
\ifx \bparticle \undefined \def \bparticle#1{#1}\fi
\ifx \barticle \undefined \def \barticle#1{#1}\fi
\bibcommenthead
\ifx \bconfdate \undefined \def \bconfdate #1{#1}\fi
\ifx \botherref \undefined \def \botherref #1{#1}\fi
\ifx \url \undefined \def \url #1{\textsf{#1}}\fi
\ifx \bchapter \undefined \def \bchapter#1{#1}\fi
\ifx \bbook \undefined \def \bbook#1{#1}\fi
\ifx \bcomment \undefined \def \bcomment#1{#1}\fi
\ifx \oauthor \undefined \def \oauthor#1{#1}\fi
\ifx \citeauthoryear \undefined \def \citeauthoryear#1{#1}\fi
\ifx \endbibitem  \undefined \def \endbibitem {}\fi
\ifx \bconflocation  \undefined \def \bconflocation#1{#1}\fi
\ifx \arxivurl  \undefined \def \arxivurl#1{\textsf{#1}}\fi
\csname PreBibitemsHook\endcsname

\theoremstyle{thmstyleone}%

\DeclareMathOperator{\arcsec}{^{\prime\prime}}

\theoremstyle{thmstyletwo}%

\theoremstyle{thmstylethree}%

\unnumbered
\begin{document}

\title[Article Title]{Evidence for the First Globular Cluster Stellar Stream beyond the Milky Way}

\author*[1]{\fnm{Julie Kiel} \sur{Holm}}\email{julie.holm@nbi.ku.dk}

\author*[1,2]{\fnm{Sarah} \sur{Pearson}}\email{sapea@dtu.dk}

\author[3]{\fnm{Jacob} \sur{Nibauer}}\email{jnibauer@princeton.edu}

\author[4]{\fnm{David J.} \sur{Sand}}\email{dsand@arizona.edu}

\author[5]{\fnm{Adrian M.} \sur{Price-Whelan}}\email{aprice-whelan@flatironinstitute.org}

\author[6,7,8]{\fnm{Tjitske} \sur{Starkenburg}}\email{tjitske.starkenburg@northwestern.edu}

\author[9]{\fnm{David} \sur{Hendel}}\email{davidahendel@gmail.com}

\author[4]{\fnm{Catherine} \sur{Fielder}}\email{cfielder@arizona.edu}

\affil*[1]{\orgdiv{DARK, Niels Bohr Institute}, \orgname{University of Copenhagen}, \orgaddress{\street{Jagtvej 155A}, \city{Copenhagen}, \postcode{2200},  \country{Denmark}}}

\affil*[2]{\orgdiv{DTU Space}, \orgname{Technical University of Denmark}, \orgaddress{\street{Elektrovej 327}, \city{Kgs. Lyngby}, \postcode{2800},  \country{Denmark}}}

\affil[3]{\orgdiv{Department of Astrophysical Sciences}, \orgname{Princeton University}, \orgaddress{\street{ 4 Ivy Ln}, \city{Princeton}, \postcode{08544}, \state{NJ}, \country{USA}}}

\affil[4]{\orgdiv{Steward Observatory}, \orgname{University of Arizona}, \orgaddress{\street{933 North Cherry Avenue}, \city{Tucson}, \postcode{85721}, \state{AZ}, \country{USA}}}

\affil[5]{\orgdiv{Center for Computational Astrophysics}, \orgname{Flatiron Institute}, \orgaddress{\street{162 5th Avenue}, \city{New York City}, \postcode{10010}, \state{NY}, \country{USA}}}

\affil[6]{\orgdiv{Center for Interdisciplinary Exploration and Research in Astrophysics (CIERA)},
\orgaddress{\street{1800 Sherman Ave}, \city{Evanston}, \postcode{60201}, \state{IL}, \country{USA}}}

\affil[7]{\orgdiv{Department of Physics and Astronomy}, \orgname{Northwestern University}, \orgaddress{\street{2145 Sheridan Rd}, \city{Evanston}, \postcode{60201}, \state{IL}, \country{USA}}}

\affil[8]{\orgname{The NSF-Simons AI Institute for the Sky (SkAI Institute)}, \orgaddress{\street{172 E. Chestnut St}, \city{Chicago}, \postcode{60611}, \state{IL}, \country{USA}}}

\affil[9]{\orgdiv{Independent Researcher},  \orgaddress{\city{Short Hills}, \postcode{07078}, \state{NJ}, \country{USA}}}

\abstract{
The dark matter content of ultra-diffuse galaxies is the subject of considerable debate \citep[][]{vanDokkum2016, Laporte2019, Kravtsov2024_DM, vanDokkum2019, Brook2021}. 
Stellar streams, which form when a host galaxy tidally strips stars from an orbiting stellar system, provide a powerful technique to constrain the dark matter content of external galaxies \citep{NibPear2025}. 
The stripped stars form long, thin leading and trailing tidal arms that persist for billions of years.
Stellar streams from globular clusters are particularly sensitive probes of dark matter halos and substructure \citep{Bovy2017,bonaca2018, Bonaca2019a, nibauer2025b}.
Globular cluster streams are expected to exist in a variety of host galaxy types \citep{Pearson2024, Le2025}, but so far, they have only been observed in the Milky Way.
We present evidence for the first extragalactic globular cluster stellar stream, identified in deep Hubble Space Telescope imaging of the ultra-diffuse galaxy, UGC9050-Dw1. 
The stream's morphology, colour, and apparent association with a compact source support the globular cluster progenitor interpretation observationally, and we reproduce the observed surface brightness with simulated globular cluster stellar populations.
We use generative stream modelling, which fits dynamical models directly to the stream morphology, to constrain the mass of the progenitor and present the first stream-based halo constraint for an ultra-diffuse galaxy.
The stream models point to a globular cluster origin and suggest a massive dark matter host halo.  
By extending the reach of globular cluster stream analysis to external galaxies,
this work opens a new chapter in dark matter science.
}

\maketitle
Recent capabilities to detect low-surface-brightness objects have led 
to the discovery of numerous ultra-diffuse galaxies (UDGs) \citep{Bothun1987, Dalcanton1997}; galaxies with dwarf-like stellar masses but Milky Way–like extents \citep{vanDokkum2015a}.
The methods available for inferring their detailed mass and density distribution remain limited, since their characteristic low surface brightness means that measuring velocity dispersions or rotation curves is challenging \citep{ForbesGannon2025}. 

\begin{figure*}[h!]
    \centering
    \includegraphics[]{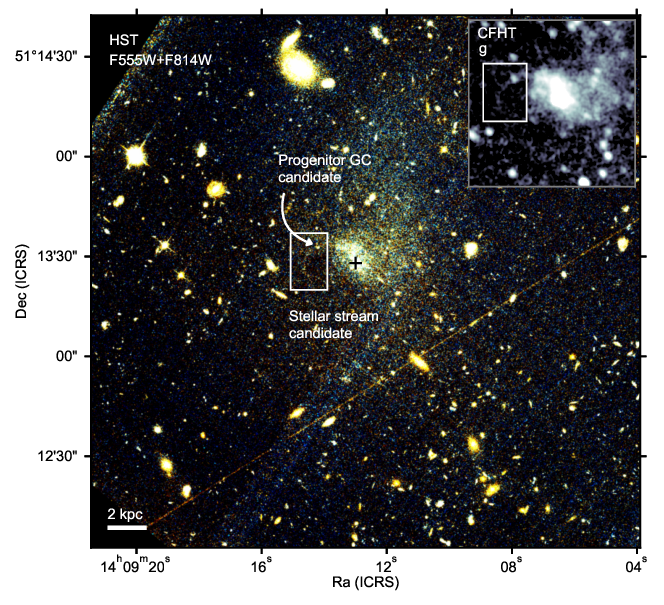}
    \caption{{\bf HST \& CFHT imaging of the candidate stream and GC progenitor in UDG9050-Dw1.} Composite image of HST ACS/WFC F555W and F814W observations of the ultra-diffuse host galaxy, UDG9050-Dw1. The \textit{Oyashio} stream and parent GC candidates are marked with a box and an arrow, respectively. The black plus marks the luminosity centre of the host galaxy as estimated by ref. \citep{Fielder2023}, and the scale bar illustrates 2 kpc at a distance of 35.2 Mpc. Insert: CFHT MegaCam $g$-band image zoomed to the vicinity of the feature.}
    \label{fig:data-images}
\end{figure*}

Figure \ref{fig:data-images} shows observations of the UDG UGC9050-Dw1 by the Hubble Space Telescope (HST) and the Canada–France–Hawaii Telescope (CFHT), in which we highlight a feature resembling one arm of a stellar stream extending from a globular cluster (GC) candidate.
The thin stream candidate and its presumed parent GC have a projected distance of $\sim2.5$ kpc to the centre of UGC9050-Dw1, which is likely associated with the low-surface brightness spiral galaxy UGC 9050 at a distance of $35.2 \pm 2.5$ Mpc \citep{Fielder2023, Mould2000}. 
UGC 9050-Dw1 was first identified in CFHT images as part of a diffuse dwarf galaxy search \citep{Bennet2017}. 
The HST image was presented in a detailed analysis of UGC 9050-Dw1's GC population, which found that most of the 
GCs likely formed simultaneously during a past dwarf merger event \citep{Fielder2023}. 
The stream-like feature, which we name \textit{Oyashio} (after a cold Pacific ocean current, pronounced [\textipa{o:.\,ja.\,\textctc i.\,o}], inspired by the nomenclature introduced by ref. \citep{Shipp2018}), is identified independently in the HST and CFHT data,
ruling out imaging or data processing artefacts as its origin. All available data are shown in Extended Data Figure \ref{fig:all-bands}.\\

A comparison between the background-subtracted image counts, $f$, of the stream candidate, and the standard deviation of the background, $\sigma$, gives a signal prominence of
$(f_{\rm stream}-\bar{f}_{\rm background})/\sigma_{\rm background}=7.34$ in the combined HST image. 
We measure the width of the stream candidate to be $w_{\pm\sigma} = 72.3\pm8.9$ pc, assuming a Gaussian profile (see \nameref{sec:methods} and Extended Data Figure \ref{fig:phi1phi2}). The amplitude, $A$, of this fit gives a signal-to-noise ratio of $A/\sqrt{\sigma_A} = 5.2$ for the HST image, and similar fits to the CFHT images give ratios ranging between 2.3 and 3.9 in all g, r, and i band images, while the feature is not detectable in the u and z bands. 

Globular cluster streams in the Milky Way (MW) have widths ranging from a few tens to a few hundred pc \citep{Bonaca2025}.
The width of \textit{Oyashio} is much smaller than any known stream from a dwarf galaxy progenitor (e.g., the \textit{Orphan-Cenab} MW stream, which has a low-mass dwarf progenitor but a width $>200$ pc \citep{Koposov2023}), pointing towards a GC origin. 
The length of the \textit{Oyashio} stream arm is 2 kpc based on a by-eye estimate. \\

\begin{figure*}[h!]
    \centering
    \includegraphics[]{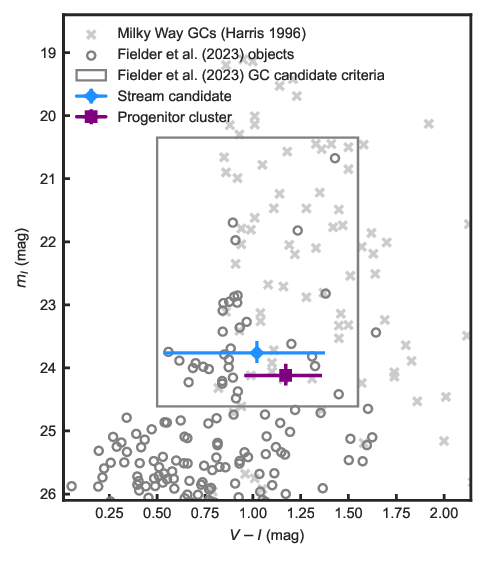}
    \caption{{\bf Colour analysis of the candidate stream and GC progenitor compared to GCs in UGC9050-Dw1 and the MW}.  The colours of the stream candidate (blue diamond) and its progenitor cluster candidate (purple square) are plotted against their magnitude in the Johnson-Cousins I band. Both colour and magnitude are integrated light. For comparison, we also plot the colours of objects studied in ref. \citep{Fielder2023} (grey circles), along with the colour and magnitude criteria they use to select GC candidates (grey box), and GCs in the MW \citep{Harris1996} (crosses). The errorbars are the magnitude uncertainties, combined with the standard deviation between the background areas, which reflect the uncertainty tied to the choice of background. The colours of the stream and cluster candidates are similar to each other, as would be expected of a GC and its stream, and they fall within the GC candidate selection box, which is anticipated for a disrupting GC in UGC9050-Dw1.}
    \label{fig:SPCMD-comparison}
\end{figure*}

The stellar population of a GC progenitor and its stream are of the same age and metallicity and therefore follow the same track (isochrone) in a colour-magnitude diagram (CMD).
In the MW these tracks can be analysed at the level of individually resolved stars. At the distance of UGC 9050-Dw1, we must consider the integrated-light equivalent: comparing the overall colour of the stream and progenitor.
Many of the GC candidates in UDG 9050-Dw1 have similar colours, likely due to simultaneous formation \citep{Fielder2023}, so we also expect this particular GC candidate and its stream to align with the colours of the other GCs in the host.
In Figure \ref{fig:SPCMD-comparison}, we show that the \textit{Oyashio} stream and progenitor cluster candidates have overlapping colours, as expected if they share the same stellar population, and both fall within the GC candidate selection box used by ref. \citep{Fielder2023}.
The GC candidate has a colour of F555W-F814W = $1.1\pm0.1$, and the stream-like feature has F555W-F814W = $1.0\pm 0.2$. This is slightly bluer than the average GC in the MW \citep{Harris1996, Larsen2001}, which aligns with the findings in \citep{Fielder2023} that the clusters in UDG9050-Dw1 are consistent with a more recent origin.

To test whether a GC stream with this morphology at this location is dynamically plausible, we apply the \texttt{X-Stream} sampler \citep{NibPear2025}, which translates stream imaging into constraints on stream progenitors and host dark matter halos.  
See \nameref{sec:methods} and the Extended Data Table 1 for details on the model parameters. 
In Figure \ref{fig:gammavsmass} (left), we show the likelihood surfaces and sampler constraints on the progenitor mass, halo mass, and on the inner density slope of the halo, $\gamma$, from the sampler applied to control points generated within a mask of width $w_{\pm2\sigma}=$ 143.7 pc.
The orange contours show the 68\% and 95\% credible regions for runs with a dark matter halo concentration of $c = 5$, and the grey contours show the same regions for the $c= 2$ runs. 

We treat these $c = 5$ runs as our fiducial runs (see posteriors in Extended Data Table 1), since varying the mask width or concentration does not make a material difference in our results, which we discuss in \ref{sec:methods}.

For our fiducial runs, the sampler places an upper limit on the initial progenitor mass of $M_{\rm prog} < 2.5 \times 10^6$ M$_{\odot}$ with 95\% confidence. 
These mass constraints are consistent with our interpretation of a GC origin for the stream \citep[][]{Bonaca2025}, and is less massive than e.g. the MW star cluster $\omega$Centauri (M$_* = 3.55\times 10^6$ M$_\odot$)\citep{Baumgardt2018} with its associated stream \citep{ibata2019}, and comparable to some GCs in M31 \citep{Usher2024}.

The surface brightness of \textit{Oyashio} in the HST images is F555W $=27.0\pm0.1$ mag/arcsec$^2$, F814W $=26.1\pm0.1$ mag/arcsec$^2$. This is brighter than the average MW stream \citep{Shipp2018}, which is to be expected for an extragalactic discovery.
We compare the measured surface brightness of the stream candidate 
to a simulated stellar population based on that of the archetypal MW GC stream \textit{Palomar 5} (Pal 5)\citep{Odenkirchen2001}, and find that a stream with the same age, metallicity and level of disruption, requires a progenitor 20 times as massive, i.e. an initial cluster mass of $\sim 2 \times 10^6$ M$_\odot$, to produce a similar surface brightness as our detected stream candidate.
We repeat this analysis for a grid of isochrones and find that a younger progenitor can produce the observed surface brightness with a progenitor mass as low as  $1.65\times10^5$ M$_\odot$ for the brightest isochrone, where we use the brightness ratio between the stream and GC candidates to gauge the level of disruption.
We note that the GC population of UGC 9050-DW1 is relatively blue and consistent with younger globular clusters than Pal 5 \citep{Fielder2023}. 

The red dots and stars in Figure \ref{fig:gammavsmass} (left) represent the lower bound and $20~ \times$ Pal 5-like progenitors, respectively.
These points are both within the 95\% confidence regions of the posteriors.

\begin{figure*}[h!]
    \centering\includegraphics[width=\textwidth]{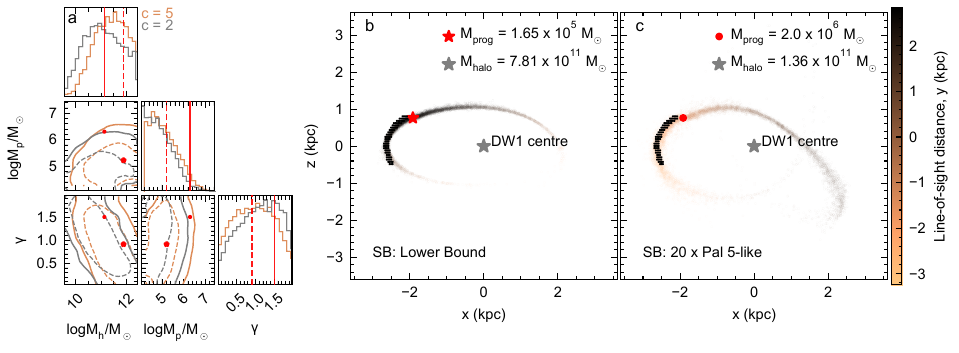}
    \caption{{\bf Dynamical constraints on the progenitor and UGC9050-Dw1 halo from the \texttt{X-Stream} sampler.} \label{fig:gammavsmass}
    {Left}: Constraint on both $M_{\rm prog}$, $M_{\rm halo}$, and the inner slope parameter, $\gamma$, from applying \texttt{X-Stream} to the GC candidate stream with a mask width of 143.7 pc (including 2$\sigma$ of the measured HST width).
    The orange and grey contours show the 68\% credible regions (dashed lines) and 95\% credible regions (solid lines) for streams evolved in a dark matter halo with $c=5$ and $c=2$, respectively. The 1D histograms show the 95\% distributions. The red star shows the lower bound progenitor mass which would lead to an observable GC stream in HST, and the red dots show the required progenitor mass to produce an observable stream for a cluster with a Pal 5-like isochrone. The red lines show the corresponding values in the 1D marginalised histograms.
    Middle: The low mass progenitor model stream generated for a fit within the 68\% credible region of the fiducial runs with $t_{\rm age} = 1100$ Myr, $\gamma =0.9$, and $\beta = 2.6$ (see red star in the full corner plot in Extended Data Figure \ref{fig: corner}). The grid of black points shows the location of the observed GC stream used as the input data for {\texttt X-Stream}. The red star indicates the progenitor position at present day, the grey star shows the centre of UGC9050-Dw1 in this galactocentric coordinate system. The stream points are coloured by their line-of-sight distances in the galacocentric rest-frame (higher values mean further from the observer). 
    {Right}: Same as middle panel but for the $20~ \times$ Pal 5-like progenitor with $t_{\rm age} = 600$ Myr, $\gamma = 1.5$, and $\beta = 3.3$. }
    \label{fig:gammavsmass}
\end{figure*} 

The \texttt{X-Stream} sampler finds a UGC9050-Dw1 halo scale mass \citep{NibPear2025} of 
log$_{10}$(M$_{\rm halo}$/M$_{\odot}) = 
11.31^{+0.67(+1.03)}_{-0.71 (-1.30)}$ within the 68\% (95\%) confidence limit, which corresponds to log$_{10}$(M$_{\rm 200}$/M$_{\odot}) = 11.63^{+0.71}_{-0.83} \; (68\%)$  using Planck cosmological parameters \citep{Planck2020}, and finds 
$\gamma= 0.92^{+0.57(+0.92)}_{-0.58(-0.87)}$ within the 68\% (95\%) confidence limits. 
This is the first constraint on a dark matter halo mass and density slope from a stellar stream in an ultra-diffuse galaxy. It suggests that this UDG is slightly less cored than the average low-surface brightness dwarf ($\gamma \sim 0.2$)\citep{deBlok2001}, and than UDG Dragonfly 44 ($\gamma =0.3$)\citep[][]{vanDokkum2019}, but allows inner slopes comparable to that of UDG AGC 242019 ($\gamma \sim 0.54$)\citep[][]{Brook2021}. 
The sampler also places strong constraints on the line-of-sight progenitor position and on the progenitor orbit, while the halo scale radius and outer density slope are unconstrained. We show the full posterior distributions for the fiducial run 
in Extended Data Figure \ref{fig: corner}. In Extended Data Figure \ref{fig:good_vs_bad}, we show examples of five streams within the  68\% confidence regions (good fits), and five outside the 95\% region (inconsistent fits). While the good fits reproduce the observed input data well, the inconsistent fits are either too straight, too curved, offset from the input data, or too wide.

In Figure \ref{fig:gammavsmass}, we show two streams models: 
The lower bound stream selected, based on our surface brightness analysis above, to have a progenitor mass of $1.65 \times 10^5$ M$_{\odot}$ (middle), and the $20~ \times$ Pal 5-like stream, selected to have a progenitor mass of $2 \times 10^6$ M$_{\odot}$ (right). All other parameters were selected at random within  
the  68\% credible region of the likelihood surface. 
Both model streams reproduce the morphology of the observed input data well, but the $20~ \times$ Pal 5-like stream (right) is longer. 
We mock-observe these two streams and show the results in Extended Data Figure \ref{fig:mock} (top and middle panel). The densest parts of the mock observed model streams look similar to the observed stream candidate, while the more diffuse end parts of the model streams are not observable.
Note that the arm towards positive $x$-values, which was not included in the fitting procedure, wraps behind the brighter central part of the UDG itself. This can help explain why we only detect one arm in the HST data.

For the model stream in Figure \ref{fig:gammavsmass} (right), for which the halo mass is at the peak of the posterior distribution, the progenitor's current galactocentric radius is $R_{\rm gal,today} = 2.52$ kpc, the inferred enclosed mass of the UDG  is M$(<R_{\rm gal,today}) = 1.36 \times 10^{10}$ M$_{\odot}$. 
The total mass (M$_{\rm 200} = 1.56 \times 10^{11}$ M$_{\odot}$) is similar to mass estimates of the Large Magellanic Cloud 
\citep{penarrubia2016, erkal2019},
and is within errors of previous halo mass estimates \citep{Fielder2023}, which used the GC count to calculate a total halo mass 
of $1.5\pm 0.3\times 10^{11}$ M$_\odot$ or $1.8\pm 0.3 \times 10^{11}$ M$_\odot$, depending on the assumptions made. 
The entire mass range of ref. \citep{Fielder2023} is within our allowed halo masses for all runs. 

The tidal radius, $r_t \propto R_{\rm gal}\left(\frac{M_{\rm prog}}{M_{\rm halo}}\right)^{1/3}$, of a globular cluster determines the boundary at which stars become unbound from the progenitor and can escape into the leading and trailing arms of the stream. If $r_t$ is much larger than the extent of the cluster, a stream cannot form (see e.g., \citep[][]{pearson2022b}). For the fit shown in Figure \ref{fig:gammavsmass} (right), the present day $r_t = 133$ pc and the minimum $r_t = 95$. 
This is similar to the estimated tidal radius of Pal 5 ($\approx 145$ pc)\citep[][]{Ibata2017}. 

Since the cluster with the lower bound progenitor mass (middle panel) is evolving in an even higher mass halo, we conclude that tidal stripping is feasible for both clusters presented in Figure \ref{fig:gammavsmass}, and that our dynamical models are consistent with a GC being tidally stripped with $M_{\rm prog}< 2.5 \times 10^6$ M$_{\odot}$ at 95\% confidence.

There are other mechanisms that could potentially create a similar structure in an image.
Recent observations have reported evidence of tidal tails from mergers of GCs or nuclear star clusters in dwarf galaxies \citep{Poulain2025}. Tidal tails are formed due to mutual interactions between two low mass ratio objects rather than from Lagrange point stripping of high mass ratio objects, such as a GC and a host galaxy halo, which leads to the formation of stellar streams. In the case of merger-induced tails, we expect them to occur near the host centre. 
This scenario is unlikely to explain the feature presented in this paper, which has a distance $>$2 kpc from the luminosity centre of the host UDG. 
To date, no extragalactic stellar streams from globular cluster progenitors have been confirmed, and previous candidates have been shown to be physically implausible \citep{Abraham2018, Pearson2019, Nibauer2023} or simply not detected at a statistically significant level above the background stellar halo, even in nearby galaxies such as M31 \citep{Pearson2022a}.

A tidal shell formed in a radial collision between two stellar systems can also appear as a thin curved structure \citep[e.g.,][]{Sanderson2013}. However, the centre of curvature for {\it Oyashio} is offset from the centre of the host, which is not expected for shells.
Similarly, a lensing effect of a background galaxy could create an arc \citep{Einstein1936, Cunha2018}, but we do not observe other arcs or any plausible lenses. Additionally, the stream's alignment with the GC candidate and the fact that their colours are similar to the other GCs in the system would need to be coincidental. Alternatively, a dusty region may give rise to fluctuations mimicking the stream candidate. If there were a dust patch in the region on one side of the stream, we would expect it to be redder than on the other side. We find no evidence of such a trend.

A stellar stream from a small dwarf galaxy might be misinterpreted as a GC stream. 
In the case presented here, the observational width measurement and the dynamical mass constraints both point to a GC progenitor.
We might only be observing the densest part of the stream, but even with a larger mask width, our modelling mass constraints point to a GC origin.   
In Extended Data Figure \ref{fig:mock} (bottom), we also mock observe a dwarf-like stellar stream with the same stellar mass as our $20~ \times$ Pal 5-like stream but including dark matter.
The dwarf-like stream has a larger surface area, yielding a lower surface brightness. For the dwarf stream to be observable, it would need a $\sim 5~ \times$ larger stellar mass than our reported upper limit of the progenitor mass, and such a stream would be observed as wider than {\it Oyashio} in the HST data.

Lastly, there is the possibility of a chance alignment of unresolved stars in the host galaxy or a chance projection of stars unassociated with the host. 
While chance alignment or chance projection cannot be entirely ruled out, the similarity in colours and the agreement between modelling results, halo mass estimates from GC counts, and our mock observations support a GC stream detection. 
It also seems likely that the first extragalactic GC stream to be detected would reside in a massive halo of a low-surface-brightness galaxy, facilitating tidal stripping while providing a faint background. 

Deeper observations with HST or JWST of this object could distinguish it further from background and thus determine its origin with greater certainty. Spectroscopic studies with e.g. the Keck telescope (similar to \citep{Haacke2025, vanDokkum2016}) could compare parts of the \textit{Oyashio} stream candidate to the presumed parent cluster to reveal similarities and differences in stellar populations. 

UGC 9050-Dw1, like many UDGs, has a complex morphology, which is one of the reasons that their dark matter mass and density profiles are not well understood \citep{Shi2021,Brook2021}. 
Stellar streams from GCs are sensitive to the local acceleration field \citep{bonaca2018,nibauer2022}, and our results provide the first-of-its-kind UDG mass and density constraints, introducing a new tool to study UDGs. 

Cold dark matter models predict the formation of low-mass subhalos devoid of stars \citep{Diemand2008}, and a detection would put tight constraints on the properties of candidate dark matter particles \citep{Bullock2017}.
A promising avenue for detecting low-mass subhalos is through their interaction with stellar streams \citep{nibauer2025b}, which can create variation in the stream density \citep{Bonaca2019a, Bovy2017}. 
Extending the sample of GC stellar streams to include extragalactic hosts, as done in this work, allows us to study host galaxies with fewer baryonic perturbers than the MW, which will increase the chances of an unambiguous subhalo detection \citep{
BanikBovy2019}. 

A GC stellar stream visible at a distance of 35.2 Mpc highlights the potential of discoveries in neighbouring diffuse galaxies, especially if GC streams are as frequent in UDGs as extrapolated MW counts suggest \citep{Pearson2024, Le2025}.
With the capabilities of the Euclid and Roman Space Telescopes \citep{Roman:Spergel2015,Racca2016}, 
we expect to detect many more GC streams
\citep{Pearson2019, Pearson2022a,Aganze2024}. 
The extragalactic globular cluster stellar stream presented in this paper serves as a precursor to future detections and represents an independent way to probe the dark matter mass and density profiles of ultra-diffuse galaxies.

\section{Methods}\label{sec:methods}
\subsection{Data}
The HST Advanced Camera for Surveys (ACS) images use the Wide Field Channel (WFC) in the F555W and F814W filters, have an angular resolution of $0.1\arcsec$ and reach exposure times of 2406 s and 2439 s, respectively.
The observations were carried out in September of 2022 under HST program ID 16890 \citep{Sand2021} and the resulting images of UGC9050-Dw1 were presented in ref. \citep{Fielder2023}. 
The images have been calibrated through the standard HST CALACS pipeline, and are available on the STScI/MAST archive. 

CFHT has observed the same region of the sky using MegaCam as part of its Legacy Survey in 2005, and UGC 9050-Dw1 was first identified and selected in a semi-automated search for diffuse dwarfs \citep{Bennet2017}. The area containing UGC9050-Dw1 is visible in both the W3-1-3 and W3-2-3 fields of the survey, which contains bands $u$, $g$, $r$, $i$, and $z$, so a total of 10 CFHT images are available.
Because of the $S/N$ levels, we are not able to confidently identify the stream-like feature in the $u$ or $z$-band images. Integration times, observation depth and seeing for the CFHT data is published in \citep{Gwyn2012}.
The CFHT data, which have been calibrated through the MegaPipe pipeline, are available through the Canadian Astronomy Data Centre.

For visual purposes, we have smoothed the HST and CFHT images using a Gaussian kernel (with $\sigma=$ 3 and 1.5 pixels, respectively), and we plot the area of interest in all available bands in Extended Data Figure \ref{fig:all-bands}. 
All data analysis is performed on unsmoothed data.

\setcounter{figure}{0}
\renewcommand{\figurename}{Extended Data Fig.}
\begin{figure*}
    \centering
    \includegraphics[width=\textwidth]{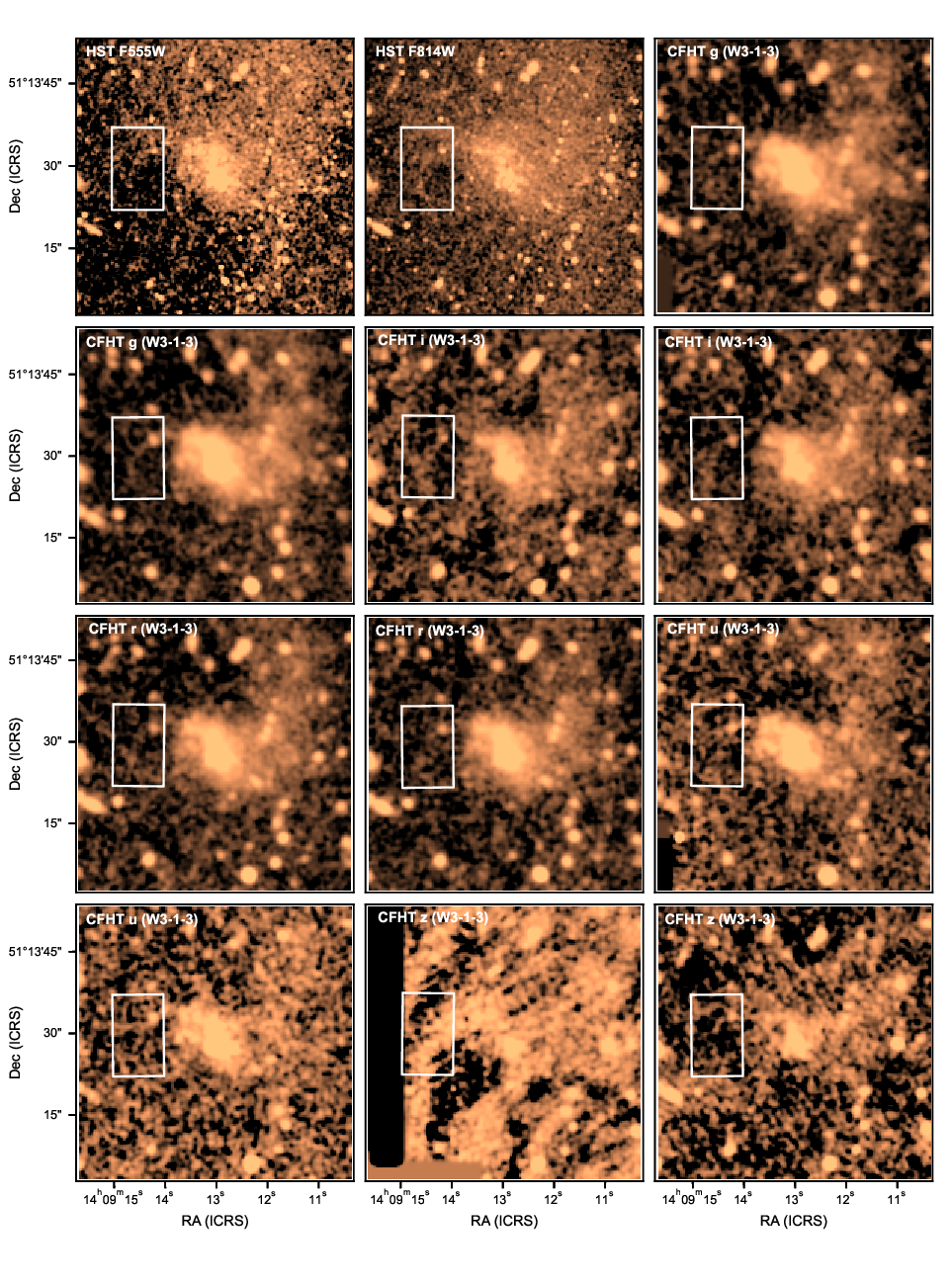}
    \caption{{\bf The area of interest in all 12 available bands.} We use a asinh stretch, a smoothing of 2 pixels for CFHT and 4 pixels for HST, and an asymmetric percentage scaling (40\% to 98\% for HST, 10\% to 98 \% for CFHT, except the z-band images, which have a lower bound of 20\%, because the invalid regions otherwise dominate the bottom centre image). The area containing the stream candidate is marked with a white box.}
    \label{fig:all-bands}
\end{figure*}

\subsection{Data analysis}
\label{sec:m-dat}
The feature was initially detected by eye, but when we apply the Rolling Hough Transform \citep{Clark2014} with a range of (hyper)parameters, the code consistently returns both the location and curvature of the stream candidate. We use this as a validation of the initial by-eye detection.

\begin{figure*}[h]
    \centering
    \includegraphics[width=\textwidth]{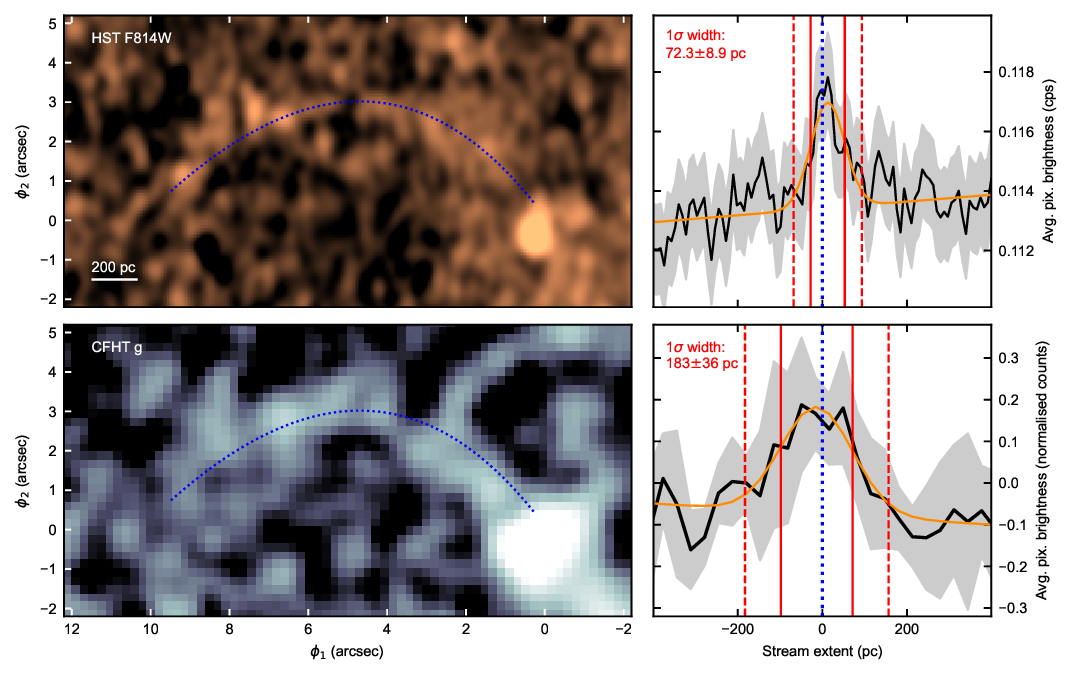}
    \caption{{\bf Width analysis of the stream candidate.} Left: HST F814W (top) and CFHT g (bottom) images rotated into a coordinate frame aligning with the stream and zoomed to display it in detail. We fit a track along the stream (blue dotted line) and calculate the average brightnesses parallel to it, following the curvature and not including the progenitor GC. Note that it is not a direct summation through $\phi_2$ values, but follows the curvature of the stream. Right: The resulting brightness average (black line) fitted along the stream track (blue) with the uncertainty on the mean shaded in grey. We perform a Gaussian fit (orange line) to the total average, which is used to estimate the width of the stream. Red lines illustrate $w_{\pm\sigma}$ (solid) and $w_{\pm2\sigma}$ (dashed), which in the HST data correspond to 72.3 and 143.7 pc, respectively. }
    \label{fig:phi1phi2}
\end{figure*}

In Extended Data Figure \ref{fig:phi1phi2} (left), we rotate the images
into a coordinate frame aligned with the stream, where $(\phi_1, \phi_2)$ represent stream longitude and latitude \citep{gala, astropy2018}. 

We draw masks around the stream by following a polynomial fit to points selected manually in the F814W image, creating a curved shape with constant width, excluding the GC candidate. We duplicate these on both sides to mask the background areas.
To measure the width, we sample slices perpendicular to the track curvature in steps of pixel size along this fit. We collapse the resulting area along the fitted track, calculating the mean brightness of the stream candidate and its surrounding areas. The right panels of Extended Data Figure \ref{fig:phi1phi2} depict the perpendicular brightness profile, consisting of the average brightnesses measured parallel to the curved stream track. This is done in six segments to check for any single dominating contaminant. We fit a Gaussian with a linear offset to the perpendicular brightness profile for each segment, as well as the whole stream. 
This way of sampling the surrounding area causes the inner part of the curve to be oversampled compared to the outer curve, and thus by construction, it has a lower error and dominates the fit. We therefore use the average of all standard deviations as uniform errors.
The standard deviation of the Gaussian provides an estimate of the width of the stream, with uncertainties on the fitted parameter. Since $\sigma$ describes the deviation from the mean, $\mu$, the total width is twice $\sigma$. We denote this width, measured from $\mu-\sigma$ to $\mu+\sigma$ , as $w_{\pm\sigma}$. The width that includes 95\% of the Gaussian area, corresponding to $2\sigma$ from the mean, we call $w_{\pm 2\sigma}$. 
The physical width is affected by the instrument point spread function (PSF), which we account for by subtracting the observational dispersion in quadrature. 
We use the full width at half maximum for HST of 0.1$\arcsec$. 

The seeing of the CFHT observation ($0.88\arcsec$) is comparable to the width of the feature in the CFHT images ($w_{\pm\sigma} = 1.1\arcsec$). Therefore, we use the HST width in subsequent analysis.

There is a brighter object located next to the track at around $\phi_1=-4$, which increases the measured width. If we only use the track up to that point, it yields a width of $w_{\pm\sigma} = 52$ pc instead. 
On the other hand, the faintness of the object means we might only be sensitive to the brightest central part of our stream candidate, which would lead to an underestimated width. In Extended Data Figure \ref{fig:mock} we illustrate that the simulated GC streams show up with similar apparent width when injected into the data. 

To quantify the prominence of the stream candidate, we sum up the counts in the on-stream mask and in the off-stream background masks. We subtract the mean background from the signal, and compare this to the standard deviation of the backgrounds:
$(f_{\rm stream}-\bar{f}_{\rm background})/\sigma_{\rm background}$. 

The Gaussian fit used in the width measurement also gives a signal-to-noise ratio when we compare the amplitude to its fit uncertainty, $A/\sqrt{\sigma_A}$. This uncertainty takes into account the error on the means, which were included in the fit. 
Neither prominence estimate includes the GC candidate; they are solely based on light from the stream candidate.

The statistical significance contains three main components: the probability that it is a stream, the probability that we would find it, and, given the first two statements are true, the statistical significance of the signal in the data. What we quote as a significance is only the latter.

\begin{figure*}[h]
    \centering
    \includegraphics[width=0.6\textwidth]{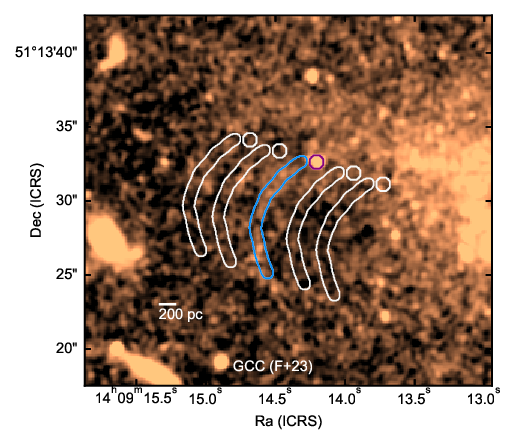}
    \caption{{\bf Masks and background regions.} Masks used for colour determination of the stream and cluster candidates (blue and purple) overplotted on the HST F814W image, along with the masks used for background subtraction (white). We use the $w_{\pm2\sigma}$ of 143.7 pc (16.8 px) for the width of the stream masks and a circle of 170.6 pc (20 px) in diameter to enclose the light of the progenitor GC. At the bottom, a GC candidate selected in a previous analysis \citep{Fielder2023} is marked for comparison.}
    \label{fig:backgrounds}
\end{figure*}

To compare the colours of the stream and cluster candidates to other GC candidates in the host UDG, we sum up their brightness in each of the two available HST filters. The stream area is defined as a curved shape with the constant width found above ($w_{\pm 2\sigma} = $16.8 pixels $=$ 143.7 pc) around a spline fitted to an array of manually selected points along the track. The cluster is contained in a circular mask 20 pixels (170.6 pc) wide, following a photometric growth-curve analysis similar to e.g. \citep{Johnson2012} to find an aperture radius which includes the majority of the light. We make sure that the masks of the stream and progenitor do not overlap. Both masks are illustrated in Extended Data Figure \ref{fig:backgrounds} with blue and purple outlines, respectively. 
All magnitudes are in the Vega system, and we correct for galactic extinction using values from \citep{SchlaflyFinkbeiner2011} as calculated through the NED NASA/IPAC extinction calculator. 

For accurate colour measurements, we need to subtract the background 
of a nearby approximately empty equal area  \citep[see e.g.,][]{Bonaca2020}. Due to the complex morphology of UGC 9050-Dw1, the choice of this area is not obvious, and the resulting colour is affected to a significant degree by the choice. We therefore use the average of several areas distributed on both sides of the stream candidate (illustrated in Extended Data Figure \ref{fig:backgrounds}). The errors on the colour values reflect the spread resulting from using either of the background areas on its own, in addition to the error on the brightness in each band, which is measured by the standard deviation of pixel values.

The ACS pixel scale of $0.05\arcsec$/pixel corresponds to 8.53 pc at this distance. This is comparable to the size expected for globular clusters ($\sim $ few pc)\citep{Brodie2006}, and ref. \citep{Fielder2023} therefore treat the GC candidates as point sources. For a partially disrupted cluster, however, we expect a larger physical extent, and by definition, a stream will not resemble a point source. We therefore treat both objects as extended sources, using areas illustrated in Extended Data Figure \ref{fig:backgrounds} as their extent. 
Our GC candidate is not included in the final GC candidate sample of ref. \citep{Fielder2023}, and inspection of their selection criteria reveals that it is discarded based on roundness, sharpness, and/or magnitude uncertainty.
A tidally disrupting GC would be both elongated and more diffuse than the undisturbed clusters, which explains the deviation from all three criteria.

We integrate the light over the areas illustrated in Extended Data Figure \ref{fig:backgrounds} to measure the surface brightness of the stream candidate. This yields a surface brightness of $27.0\pm0.1$ and $26.1\pm0.1$ mag/arcsec$^2$  in the F555W and F814W filters, respectively, after subtracting the background. 

To explore the possible stellar populations in the stream, we use a grid of isochrones with varying age and metallicity. We sample stars from a simple initial mass function (IMF) following $\frac{dN}{dm}\propto m^{-0.5}$ \citep{Grillmair2001} and truncated based on the mass range of survived stars in each of the isochrones. We extract brightnesses from the interpolated isochrone tracks, generated for HST and CFHT filters through the CMD input form \citep{cmdinput} using the PARSEC 1.2S evolutionary model \citep{Chen2015}. We distribute these stars across the area of the stream in the data, including the presumed hidden arm, and we calculate their surface brightness. 
We decrease the population size until there is only one isochrone that gives a surface brightness comparable to that in the HST data. This population is just visible with the brightest isochrone in our suite, and we use its total mass as a lower bound on the possible size of a stream with this surface area, producing the observed brightness.

We find that an initial progenitor mass as low as $ M_{\star, \rm tot} = 1.65 \times 10^5 M_\odot$, with an 8 Gyr old stellar population with $[M/H]=-1.0$ can reproduce the observed surface brightness of the present day stream. 

Since we do not know the detailed mass loss history of the GC candidate, in order to get the total mass of the stream and progenitor, we assume that the brightness ratio between the cluster and stream is an appropriate proxy for the mass ratio. Assuming that the two stream arms are identical, this gives a ratio of M$_{\rm stream}/$M$_{\rm total} \sim 0.75$, which we use to scale the inferred stream mass to a total progenitor mass. 

For the Pal 5-like stream, we use the same methodology to estimate the level of disruption as in ref. \citep{Pearson2019}. We use an isochrone track with age $11.3$ Gyr and metallicity of $ [M/H]= -1.3$. Pal 5 contains 3000 stars in the DECAM $g$ band between magnitudes 23 and 20 \citep{Bonaca2020}, so we sample the IMF until this threshold is reached, and use the resulting sample to calculate the surface brightness of the population. We repeat this process for samples of 5, 10 and 20 times the threshold (corresponding to equal multiplications of Pal 5 mass), the last of which yields a surface brightness matching the values from the HST data. This method assumes a mass loss identical to that of Pal 5.

\subsection{Generative stream models}
To generate stream models, we apply the \texttt{X-Stream} sampler \citep{NibPear2025}, which uses the particle-spray method \citep{fardal2015} implemented in the GPU code \texttt{streamsculptor} \citep{nibauer2025}. 
We fix the escape conditions to those of ref. \citep{fardal2015}. These particle-spray simulations are tuned to reproduce more expensive N-body simulations of globular cluster dissolution. Note that our results are dependent on the stream model escape conditions. 
We generate streams in a halo potential meant to represent the diffuse dwarf with an inner slope, $\gamma$, and outer slope, $\beta$ (see Eq. 1 in ref. \citep[][]{NibPear2025}). We do not include the baryons in our modelling efforts, as these make up an insignificant fraction of mass compared to the estimated mass of the total system (M$_* \sim 10^7$ M$_{\odot}$ within the half light radius)\citep{Fielder2023}. The dark matter halo profile is thus treated as a representation of the entire diffuse dwarf. 

\renewcommand{\tablename}{Extended Data Table}
\begin{table}[h]
\caption{Summary of the dynamical modelling. The prior range, posteriors, and confidence intervals for all 10 free parameters, as well as the values of the fixed and input parameters. }\label{tab:freeparams}%
\begin{tabular}{@{}lllll@{}}
\toprule
 {\bf Fitted parameters}    & priors & posteriors$^a$ & 68\%(95\%)    & unit\\
\midrule
$y_{\rm prog}$ &   [$-15$, 15] & $-0.2$ &$^{+3.5(+7.2)}_{-3.3 (-7.2)}$ &   [kpc]   \\
$v$  & [10, 200] & $120$ &$^{+51(+72)}_{-53(-86)}$ & [km/s]\\
$v_{\theta}$  & [0, $\pi$/2] & $1.1$ &$^{+0.1(+0.3)}_{-0.2(-0.3)}$ &  \\
$v_{\phi}$&[0, $\pi$]& $0.4$ &$^{+0.4(+0.6)}_{-0.3(-0.3)}$ &\\
log$_{10}{\rm M}_{\rm prog}$&[4.0, 7.5] & $<6.4$ & at 95\% &[log$_{10}$M$_{\odot}$]\\
log$_{10}{\rm M}_{\rm halo}$&[9.5, 12.5]& $11.3$ &$^{+0.7(+1.0)}_{-0.7 (-1.3)}$ &[log$_{10}$M$_{\odot}$]\\
$r_s$   &  [0.5, 10]  & $5.7$ &$^{+2.7(+3.9)}_{-2.6(-4.1)}$ &[kpc]  \\
$\gamma$& [0, 2]&  0.9 & $^{+0.6(+0.9)}_{-0.6 (-0.9)}$ & \\
$\beta$ &[2, 4]& $2.9$ &$^{+0.7(+1.0)}_{-0.7(-0.9)}$ &  \\
$t_{\rm age}$ &  [0.5, 3]& $1.38$ &$^{+0.69(+1.00)}_{-0.69(-1.02)}$  &[Gyr]\\
\botrule
 {\bf  Fixed parameters }  &  & & \\
\midrule
$x_{\rm prog}$ & $-1.99$ & && [kpc]  \\
$z_{\rm prog}$ & 0.54 &&& [kpc]   \\
$c$ & 2 or 5 && \\
\botrule
 {\bf  Input parameters}  &&   & \\
\midrule
mask width  & $143.7$ or 185  &&  &[pc]  \\
\botrule
\end{tabular}
$^a$For the \texttt{X-Stream} fiducial run with $c=5$ and a mask width of 143.7 pc.
\end{table}

We define the coordinate system in which we generate model streams by transforming the candidate GC progenitor position and the centre of UGC 9050-Dw1 from Right Ascension and Declination to a galactocentric rest-frame where UGC 9050-Dw1 is at the origin using 
\texttt{astropy} \citep{astropy2018} and the distance to UGC 9050-Dw1, 35.2 Mpc. $x$ and $z$ defines the plane of the sky, and $y$ is the line-of-sight direction pointing towards the observer. We fix the on-sky position of the progenitor ($x,z$) in our models.   

In this coordinate system, we outline a mask for the observed stream extending to the left of the candidate GC progenitor. 
To test how the choice of width affects our results, we use two different mask widths to represent the stream. Our fiducial mask width is $w_{\pm2\sigma}=$ 143.7 pc. We also test a mask width of 185 pc corresponding to $w_{\pm3\sigma}$.
We create a grid of control points within the outlined masks to represent the candidate stream. These control points are used as our input for the \texttt{X-Stream} sampler. Note that we only generate control points for the left side of the stream, as we do not confidently observe a right extension of the stream. 

From the control points, we generate a kernel density estimation (KDE). For each model stream, the sampler generates a KDE and uses a KL-divergence test \citep{kullback1951, Sanderson2015} to measure the difference between the KDE from each generated model stream and the KDE from the input control points (see details in ref. \citep[][]{NibPear2025}).
The best fitting model streams are ones that minimise this difference. Note that particle-spray models do not reproduce the detailed density along the stream, but that \texttt{X-Stream} relies on control points and includes an inverse density weighing to smooth over features. Streams wider than the data have excess density inconsistent with the control points, and \texttt{X-Stream} thereby disfavours wider models.

We draw a bounding box around the stream mask to ensure that we do not penalise our generated model streams if they are longer than the mask, nor if they extend to the right of the progenitor. However, the model streams are penalised if they are shorter than the observed stream. 

\begin{figure*}
    \centering\includegraphics[width=\textwidth]{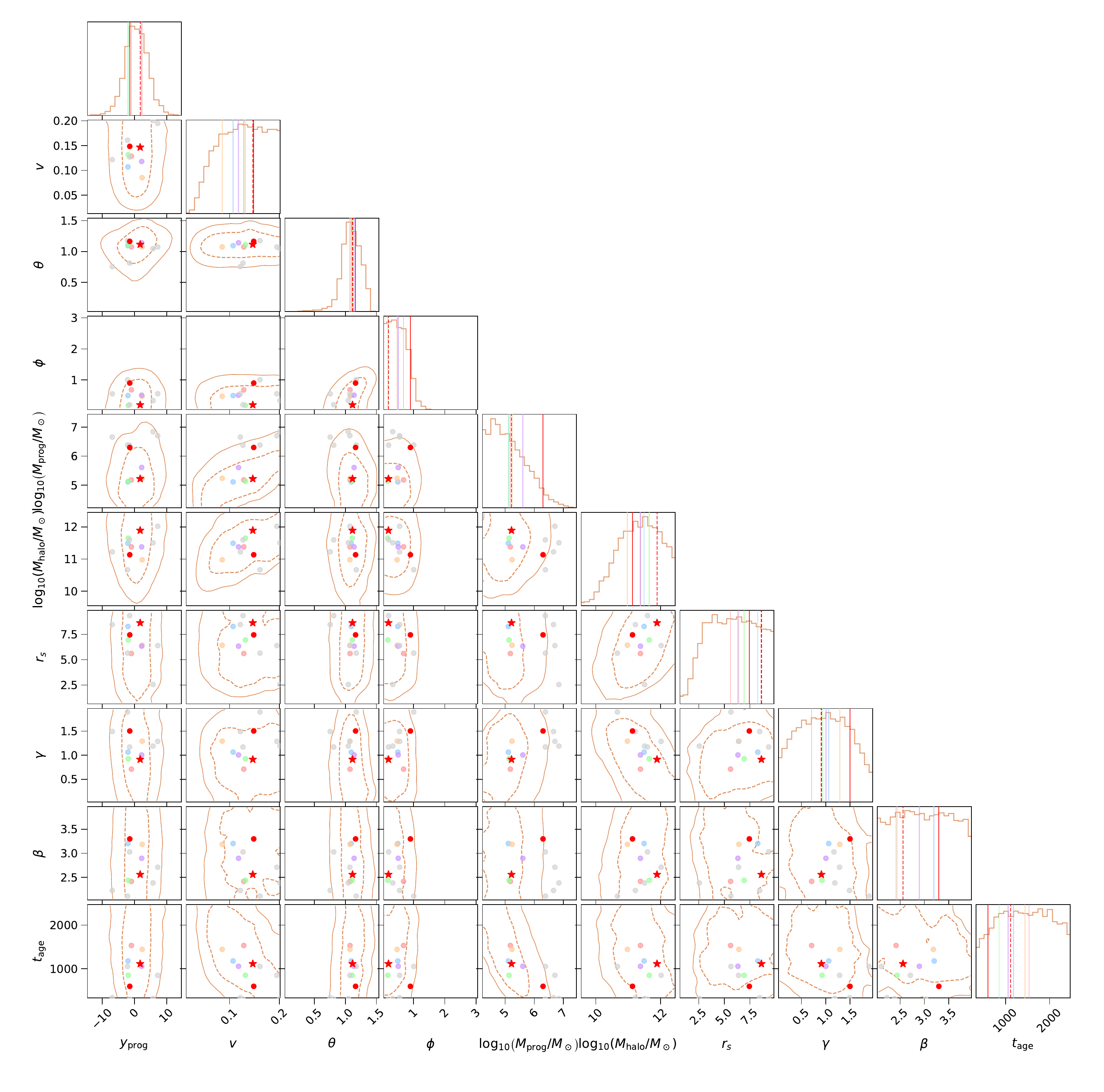}
    \caption{{\bf Full corner plot showing all constraints from the \texttt{X-Stream} sampler}. Constraints on the 10 free parameters for the GC stream candidate. The inner contour shows the 68\% credible regions, and the outer contour shows the 95\% credible regions. 
    The sampler places constraints on the line-of-sight progenitor position ($y_{prog}$), the progenitor mass, the progenitor velocity, $\theta$, and halo mass at the 68\% level. The sampler also places lower limits for the velocity magnitude, and the inner slope of the density profile, and places upper limits on the progenitor velocity, $\phi$, and $t_{age}$. There are no limits on the scale radius or the outer slope parameter. 
    The red dots and red stars show examples of the two fits within the 68\% credible region for the lower progenitor mass bound and $20\times$ Pal 5 mass progenitor stream models visualised in Figure \ref{fig:gammavsmass} (right). The pastel dots show examples of five other 1$\sigma$ consistent fits selected at random in all 10 dimensions. The grey dots show examples of inconsistent ($>2\sigma$ outliers) stream models. The model streams with these parameters are shown in Extended Data Figure \ref{fig:good_vs_bad}.
    The vertical lines show the values of each fit in the 1D marginalised histograms.}
    \label{fig: corner}
\end{figure*} 

To sample the parameter space and test which orbits, progenitor properties, and halo properties can reproduce the present-day observed candidate stream, we use 10 free parameters with uniform, flat priors and fix only the progenitor position in the plane of the sky (see Extended Data Table 1). We test two different concentrations for the dark matter halo: $c = 2$ and $c = 5$ (as motivated from \citep[][]{Brook2021, Kravtsov2024_DM}). {\texttt X-Stream} uses a nested sampling technique to probe the large parameter space, and the sampler outputs the likelihood of each model stream and the likelihood surface, including the 68\% and 95\% credible regions for all parameters. 

\begin{figure*}
    \centering
    \includegraphics[width = \textwidth]{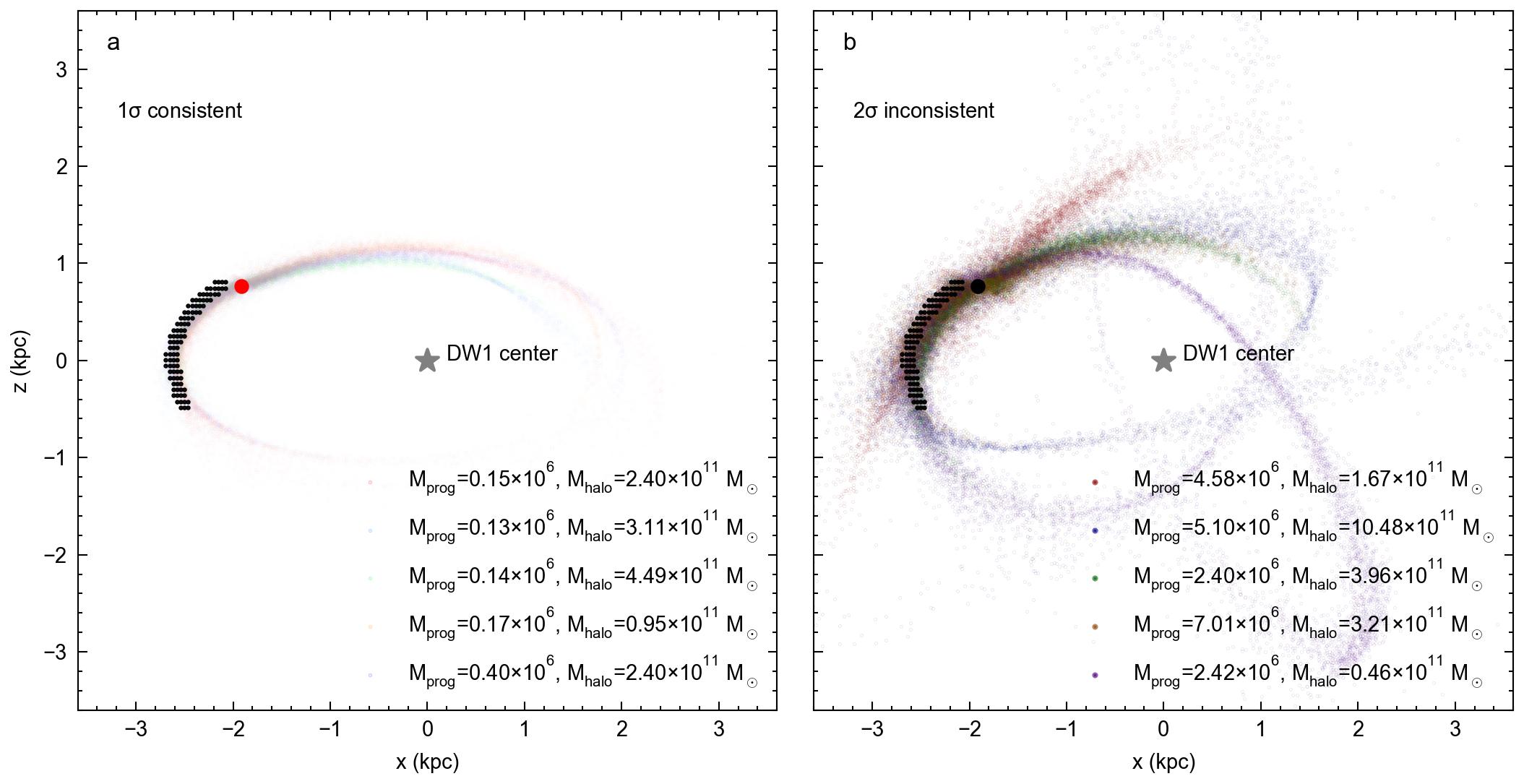}
    \caption{{\bf Visualisation of consistent and inconsistent stream models}. Left: five examples of different stream models which are consistent with the observed data. These model streams were selected to be within $1\sigma$ at random in all 10 dimensions of the posteriors in Extended Data Figure \ref{fig: corner}. Right: five examples of inconsistent stream models selected at random but 
with three or more parameters falling $>2\sigma$ outside the posterior distributions.}
    \label{fig:good_vs_bad}
\end{figure*} 

Due to the known degeneracies for the orbital direction of a stellar stream in the plane of the sky (see figs. 10 and 8 in refs \citep[][]{pearson2022b, NibPear2025}, respectively), we only sample half of the unit sphere in velocity space. 
Thus, there are islands of solutions for which the model streams move in the opposite direction (i.e. opposite leading and trailing arm), which are equally good fits. However, the distance gradient for such solutions remains the same (see figs. 10 and 9 in refs. \citep[][]{pearson2022b, NibPear2025}, respectively). 
We list all free and fixed parameters for the \texttt{X-Stream} sampler in Extended Data Table 1. See \citep{NibPear2025} for a detailed description of each parameter.

To test which parts of the model streams would be observable in the HST data, we mock-observe three different stellar stream models: 

1) The lower bound stream: A stream with a progenitor mass of $1.65 \times 10^5$ M$_{\odot}$ with [M/H] = $-1.0$ and age 8 Gyr motivated from the surface brightness analysis, which showed that such a stream would be observable in the HST data. We populate this stream with $2.8\times 10^{5}$ stars corresponding to the simulated population
with $1.65 \times 10^5$ M$_{\odot}$ described above.

2) The $20~ \times$ Pal 5-like stream: A stream with a progenitor mass of $2 \times 10^6$ M$_{\odot}$ with [M/H] = $-1.3$ and age 11.3 Gyr, which should also be observable in the HST data based on our surface brightness analysis. We populate this stream with $1.17\times 10^{6}$ stars, again based on the count in our simulated populations. 

3) A dwarf-like stream: A stream with a progenitor mass of $2 \times 10^7$ M$_{\odot}$, to mimic a more massive dwarf stream. We use the same stellar population as for the $20~ \times$ Pal 5-like stream to allow for easy comparison and to 
reflect that the extra mass is in dark matter. 

For each stream model, we sample the IMF up to the stellar mass that survives at present day, and assign
each star in the population a magnitude in the F555W and F814W HST bands, interpolated from the chosen isochrone.
We add up the distributed flux from all the stars in the model stream, add Poisson noise, and inject the mock stream into the HST image in both bands, convolving the magnitude with an Empirical PSF described by \citep{Revalski2024}, which uses PSF models from \citep{Anderson2016}. We inject each stream in approximately empty areas near the {\it Oyashio} stream to facilitate visual comparison. 

\begin{figure*}
    \centering\includegraphics{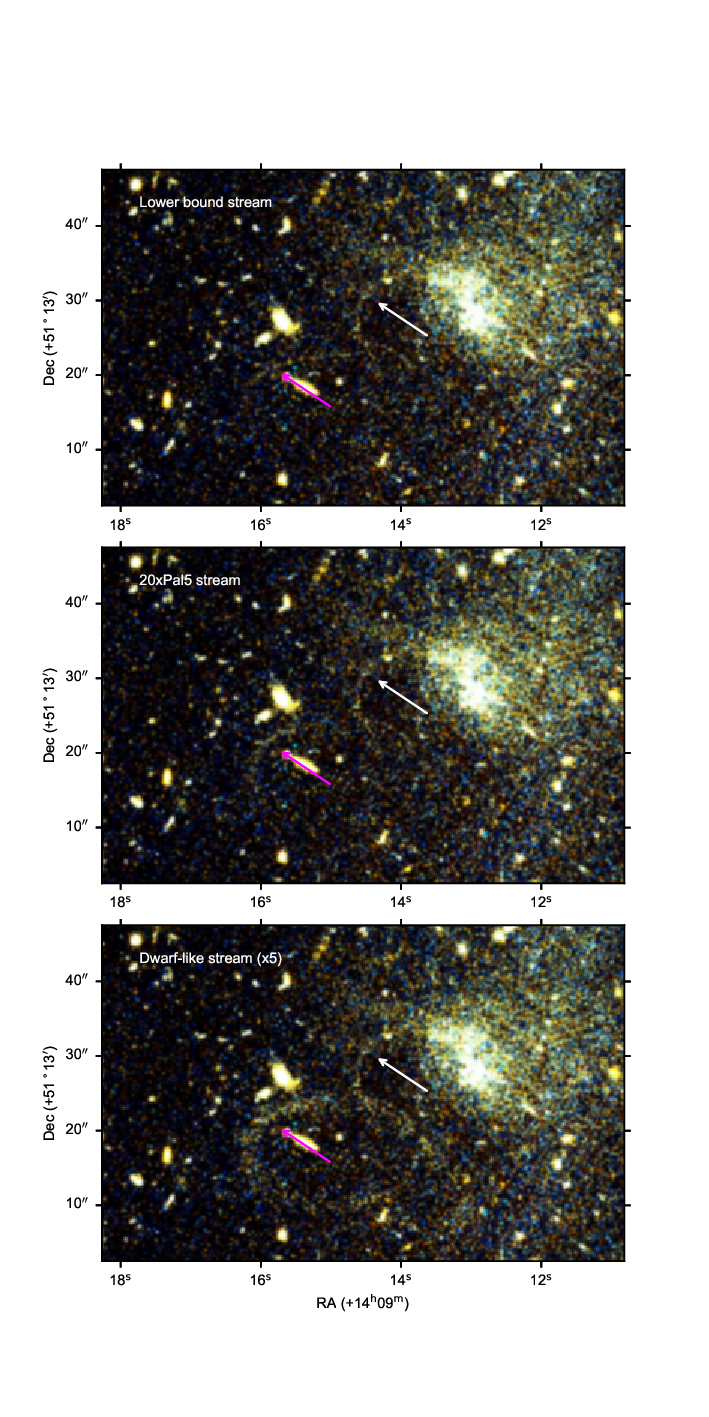}
    \caption{\textbf{Mock observations of three model streams.} The lower bound stream (top), the $20~ \times$ Pal 5-like stream (middle), and the dwarf-like stream (bottom). In all panels, the white arrow points to the detected {\it Oyashio} stream, and the magenta arrow to the injected mocks. Note that the dwarf-like stream surface brightness has been enhanced by a factor of 5 to account for the increased surface area. }
    \label{fig:mock}
\end{figure*}

When calculating the expected surface brightness of the simulated stellar population, we assume that all the stars are within the mask, which is not the case, especially for the broader dwarf-like stream. To make the dwarf-like stream visible, we multiply the brightness by 5. This illustrates that a stream with a larger area needs a more massive stellar component to achieve the same surface brightness. 
The images are stacked as in Figure \ref{fig:data-images} and shown in Extended Data Figure \ref{fig:mock}.

Note that several assumptions go into this analysis, such as the specific isochrones, degree of disruption, and the details of the stream modelling. Nonetheless, these mock observations provide a zeroth-order test of which parts of each model stream could be observable in the HST data. \\ 

\noindent \textbf{Acknowledgements.}
This version of the article has been accepted for publication, after peer review, but is not the Version of Record and does not reflect post-acceptance improvements, or any corrections. The Version of Record is available online at: http://dx.doi.org/10.1038/s41586-026-10878-w.
The Tycho supercomputer hosted at the SCIENCE HPC Center at the University of Copenhagen was used to support this work.  \\

\noindent \textbf{Funding Statements.}
S.P. and J.K.H. disclose support for the research and publication of this work from Villum Fonden (grant number VIL53081). S.P. acknowledges support from the European Union (ERC, BeyondSTREAMS, 101115754). Views and opinions expressed are, however, those of the author(s) only and do not necessarily reflect those of the European Union or the European Research Council. Neither the European Union nor the granting authority can be held responsible for them.
D.J.S. acknowledges support from NSF grants AST-2205863 and 2508746.
T.S. gratefully acknowledges support from NSF through grant AST-2510183 and by NASA through grants 22-ROMAN22-0055 and 22-ROMAN22-0013.\\

\noindent \textbf{Author Contributions.}
J.K.H. led and performed the data analysis and wrote the manuscript. 
S.P. led and performed the modelling efforts and co-wrote the manuscript. 
J.N. provided significant insight to the modelling efforts. 
D.J.S. provided significant insight to the data analysis. 
C.F. provided insight to the data analysis and additional data products. 
A.P-W. provided insight on the statistical analysis of the feature. 
T.S. provided insight to the data analysis.
D.H. had the initial idea for the project and provided insight to the data analysis. 
All authors provided feedback on the work. \\

\noindent \textbf{Competing Interests} The authors declare no competing interests.\\

\noindent \textbf{Supplementary Information.} No Supplementary Information is available for this paper.\\

\noindent \textbf{Correspondence.} Correspondence and requests for materials should be addressed to Julie Kiel Holm.\\

\noindent \textbf{Reprints and permissions.} Reprints and permissions information is available at www.nature.com/reprints.\\

\noindent \textbf{Data Availability}
The HST data are available on the STScI/MAST archive under HST program ID 16890. 
The CFHT data are available through the Canadian Astronomy Data Centre under the observation ID MegaPipe.267.282.\\

\noindent \textbf{Code Availability}
This work made use of the publicly available software: astropy \citep{astropy2018}, gala \citep{gala,adrian_price_whelan_2020_4159870},  streamsculptor \citep{nibauer2025}, and JAX \citep{Jax2018}. All code is available in the public GitHub repository juliekh/extragalacticGCstream. The simulations are based on the X-Stream code \citep{NibPear2025}, which can be made available upon request. \\

\end{document}